\documentclass[final,3p,times,twocolumn]{elsarticle}
\usepackage{graphics}
\usepackage{amssymb}
\journal{Journal of Luminescence}

\def\T2{$\mathrm{T_2}$}
\def\Pr3{$\mathrm{Pr^{3+}}$}
\def\ket#1{$\left|#1\right>$}
\def\mket#1{\left|#1\right>}

\def\PrYSO{Pr$^{3+}$:Y$_2$SiO$_5\,$}

\begin{document}

\begin{frontmatter}


\title{Towards an efficient atomic frequency comb quantum memory}

\author{A.~Amari\corref{cor1}}
\ead{atia.amari@fysik.lth.se}
\cortext[cor1]{Corresponding author}
\author{A.~Walther\corref{cor2}}
\author{M.~Sabooni}
\author{M.~Huang}
\author{S.~Kr\"{o}ll}
\address{Department of Physics, Lund University, P.O.~Box 118, SE-22100 Lund, Sweden}
\author{M.~Afzelius\corref{cor2}}
\author{I.~Usmani\corref{cor2}}
\author{B.~Lauritzen\corref{cor2}}
\author{N.~Sangouard\corref{cor2}}
\author{H.~de Riedmatten\corref{cor2}}
\author{N.~ Gisin\corref{cor2}}
\address{Group of Applied Physics, University of Geneva, CH-1211 Geneva 4, Switzerland}

\date{\today }

\begin{abstract}
We present an efficient photon-echo experiment based on atomic frequency combs [Phys. Rev. A 79, 052329 (2009)]. Echoes containing an energy of up to 35\% of that of the input pulse are observed in a Pr$^{3+}$-doped Y$_2$SiO$_5$ crystal. This material allows for the precise spectral holeburning needed to make a sharp and highly absorbing comb structure. We compare our results with a simple theoretical model with satisfactory agreement. Our results show that atomic frequency combs has the potential for high-efficiency storage of single photons as required in future long-distance communication based on quantum repeaters.
\end{abstract}

\end{frontmatter}

\maketitle

\section{Introduction}
The distribution of entanglement over long distances is a critical capability for future long-distance quantum communication (e.g. quantum cryptography) and more generally for quantum networks. This can be achieved via so-called quantum repeaters \cite{Briegel1998,Duan2001,Simon2007,Sangouard2009}, which can overcome the exponential transmission losses in optical fiber networks. Quantum memories (QM) for photons \cite{Julsgaard2004a,Chaneliere2005,Eisaman2005,Choi2008,Riedmatten2008} are key components in quantum repeaters, because the distribution of entanglement using photons is of probabilistic nature due to the transmission losses over long quantum channels. QMs enables storage of entanglement in one repeater segment until entanglement has also been established in the adjacent sections. For quantum repeaters a QM should be able to store single-photon states with high conditional fidelity $F$ and with high storage and retrieval efficiency, $\eta$ \cite{Sangouard2009}. Further, it has recently been shown that in order to reach useful entanglement distribution rates in a repeater, QMs with multiplexing capacity (multimode QM) are necessary \cite{Simon2007,Collins2007}.

	Significant progress have been achieved lately using atomic ensembles for manipulating the propagation and quantum state of an optical field, see Hammerer et al. \cite{Hammerer2008} for a recent review. Storage of single photons using electromagnetically induced transparency (EIT) has been demonstrated with warm \cite{Eisaman2005} and cold vapors \cite{Chaneliere2005,Choi2008} of alkali atoms. Storage of light at the single photon level has been demonstrated also in rare-earth-ion-doped crystals (REIC) \cite{Riedmatten2008}. REICs are characterized by large optical inhomogeneous broadening which enables storage and recall of coherent information by manipulating and controlling the inhomogeneous dephasing using echo techniques. Although traditional photon echoes cannot readily be used in the single-photon regime due to spontaneous-emission noise induced by the $\pi$-pulse \cite{Ruggiero2009}, photon echo techniques avoiding this noise have been proposed; controlled reversible inhomogeneous broadening (CRIB) \cite{Moiseev2001, Nilsson2005a, Kraus2006, Alexander2006,Tittel2008} and more recently atomic frequency combs (AFC) \cite{Afzelius2009a}. The AFC protocol may offer a breakthrough for the practical construction of quantum repeaters capable of achieving sufficient entanglement distribution rate, since the number of modes that can be stored in an AFC QM is independent of the memory material absorption depth. Since the proposal of the AFC scheme, storage of light pulses at the single photon level (so called weak coherent states) has been shown in Nd$^{3+}$:YVO$_4$ \cite{Riedmatten2008}. The fidelity of the storage was measured by storing a time-bin qubit and performing an interference measurement on the recalled qubit. The resulting interference fringe visibility was $V$=95\%, which corresponds to a fidelity $F=(1+V)/2$ \cite{Staudt2007a} of 97.5\%. This shows that light at the single photon level can be stored and retrieved without introducing noise, and future experiments are likely to improve the fidelity further. The combined storage and retrieval efficiency, however, was only 0.5\% in that experiment. A more recent experiment in Tm$^{3+}$:YAG \cite{Chaneliere2009} showed improved efficiency of 9\%, also with weak coherent states. Finally, storage of 64 weak coherent states encoded in different temporal modes has been achieved in Nd$^{3+}$:Y$_2$SiO$_5$ \cite{Usmani2009}, underlying the high multimode capacity of the AFC scheme. In view of these encouraging results in terms of fidelity and multimode storage, it is clear that increasing the efficiency is of great importance, particularly in the perspective of future long-distance quantum repeaters where QM efficiencies of 90\% are necessary with the architectures known today \cite{Sangouard2009}.
	
	Here we report a photon-echo experiment based on an AFC in a Pr$^{3+}$-doped Y$_2$SiO$_5$ crystal. We measure echoes containing an energy of up to 35\% of that of the input pulse, which is the highest AFC echo efficiencies measured so far. This shows that AFC-based schemes can be used for efficient light storage. This improvement is possible because of a good control of the procedure that creates the atomic frequency comb, via optical pumping techniques, and because of a storage medium with high optical depth. Combs with peaks of widths 100-300 kHz with peak absorption depths approaching 10 were created inside a transparent region created by optical pumping techniques in a part of the inhomogeneous profile in a \PrYSO crystal. We also examine parameters related to the experimental optimization of the efficiency and compare to a theoretical model in order to understand how to further improve the efficiency of storage and retrieval from a memory using the AFC scheme.

The paper is organized in the following way. In Sec. \ref{theory} we give an overview of the theory of AFC. In Sec. \ref{exp} the experimental setup is described. Preparation of a narrow periodic series of absorption peaks is discussed in Sec. \ref{Preparation}. In Sec. \ref{result} experimental results of AFC echoes are presented and compared with the theoretical model. Conclusions are given in Sec. \ref{conclus}.

	
	\section{Theory of AFC}
	\label{theory}

We consider an ensemble of atoms with a transition $|g\rangle-|e\rangle$ having a narrow homogeneous linewidth $\gamma_h$, but a large inhomogeneous broadening $\Gamma_{in} \gg \gamma_h$. There are thus many addressable spectral channels within the optical transition. We also assume that there is at least one more meta-stable ground state, $|aux\rangle$, having a long population lifetime. This allows a high-resolution spectral shaping of the $|g\rangle - |e\rangle$ transition by spectral hole burning, where $|aux\rangle$ is used as population storage reservoir. These properties are often found in rare-earth-ion-doped crystals \cite{Macfarlane1987,Macfarlane2002}, which are considered here. The detailed experimental procedure for precise spectral shaping depend on the particular system. In Section~\ref{Preparation} we discuss the procedure for Praseodymium doped Y$_2$SiO$_5$ crystals.

We assume that the inhomogeneously broadened transition has been shaped into a periodic series of narrow peaks, called an atomic frequency comb, see Fig \ref{AFC_comb}. We further assume that the light pulse to be stored has a spectral bandwidth, $\gamma_p$, larger than the periodicity in the comb ($\gamma_p>\Delta$), but smaller than the total comb structure. The interaction between the input pulse and a ground-state population grating versus frequency generally results in a photon echo emission after a time $1/\Delta$, which is used in accumulated or spectrally programmed photon echoes \cite{Hesselink1979,Carlson1984,Mitsunaga1991a,Yano1992,Merkel1996b,Tian2001a}. The echo emission arises from the evolution of the atomic coherence induced by the input pulse, which periodically rephases due to the periodicity in the atomic population grating. In typical echo experiments only a small fraction of the input pulse is re-emitted in the echo and the storage time is not variable since it is set by the predetermined grating periodicity $\Delta$. This is not useful for quantum repeaters where efficiencies close to 100\% and on-demand read-out of the quantum memory is necessary \cite{Simon2007}. Solutions to these issues were, however, recently proposed in Ref. \cite{Afzelius2009a}.

In Ref. \cite{Afzelius2009a} it is shown theoretically that a comb-shaped grating consisting of sharp and strongly absorbing peaks could generate a very efficient echo. This can be understood in terms of the Fourier-transform of the grating function, which governs the evolution of the atomic coherence. The periodicity in frequency results in a periodic time evolution, with an overall decay given by the width of the peak in the comb. For a series of well-separated Gaussian peaks, with full width at half maximum (FWHM) $\gamma$, the decay (dephasing) is given by $e^{-t^2\tilde{\gamma}^2/2}$ where $\tilde{\gamma}=\gamma/ \sqrt{8 \ln 2} $. For the first echo emission at $t=2\pi/\Delta$ this dephasing factor becomes $e^{-\frac{1}{F^2} \frac{\pi^2}{4 \ln 2}}$ (note that the factor applies to the field amplitude), where $F=\Delta/\gamma$. From this observation it follows that a high-finesse grating strongly reduces the intrinsic dephasing. In general the dephasing factor (for the field amplitude) is given by the Fourier-transform of one peak in the comb.

Obtaining a high efficiency echo also requires a strong interaction between the ensemble of atoms and the field, which can be achieved by a high absorption depth, $d$. It is shown in Ref. \cite{Afzelius2009a} that the comb absorbs uniformly over the photon bandwidth, under the assumption that $\gamma_p>\Delta$. The effective absorption $\tilde{d}$ depth depend on the exact shape of the peaks in the comb, but in general it decreases with increasing $F$ for a given peak amplitude, since the total number of atoms decreases. For Gaussian peaks one finds that $\tilde{d} \approx d/F$, and the fraction of the input light that is transmitted through the AFC is given by \cite{Afzelius2009a}

\begin{eqnarray}
T =e^{-\tilde{d}},
\end{eqnarray}

\noindent while the absorption is simply given by $1-T$. For an AFC consisting of peaks with Gaussian line shape the resulting echo efficiency is given by (see \cite{Afzelius2009a} for the derivation)

\begin{eqnarray}
\eta = \tilde{d}^2 e^{-\tilde{d}} e^{-\frac{1}{F^2} \frac{\pi^2}{4 \ln 2}},
\end{eqnarray}

\noindent where qualitatively the first factor can be understood as the coherent response of the sample, the second factor the re-absorption of the echo and the last factor the previously mentioned dephasing. For a high finesse, $F$, and high peak absorption $d$, the efficiency tends to a maximum of 54\% for an effective absorption depth $\tilde{d}$=2, limited by re-absorption of the echo. Higher efficiency can be achieved using three-level storage and counter-propagating fields \cite{Afzelius2009a} (see below). In this work we show experimental efficiencies up to 35\%, which is significantly higher than previous AFC experiments. This improvement results from our ability to make high finesse, high absorbing comb structures.

We also note that a solution to the predetermined storage time was proposed in \cite{Afzelius2009a} (see above). It is based on coherent transfer of the excited state amplitude to a long-lived ground state coherence, for instance a spin coherence, before the appearance of the echo. The memory can be read-out by transferring back the amplitude to the excited state, after a time determined by the user. This aspect of the proposal was recently demonstrated experimentally \cite{Afzelius2009b}. A three-level system and counterpropagating control pulses allows for a spatial reversal of the propagation of the echo (so called backward recall). In the absence of dephasing backward recall can reach 100\% efficiency by cancelation of the re-absorption, as discussed in Refs. \cite{Moiseev2001,Nilsson2005, Kraus2006,Sangouard2007,Afzelius2009a}

\section{Experiment}
\label{exp}

The measurements were performed on the site 1 transition $^1D_2-^3H_4$ at 605.977 nm in a \PrYSO crystal immersed in liquid helium at a temperature close to 2.1 K. The sample was $20\times10\times10$ mm$^3$ and had a Pr$^{3+}$ concentration of $0.05\%$ which gives an absorption depth in the range $60<d<80$ \cite{Walther2009a} at the center of the inhomogeneous profile. The high absorption was critical in order to obtain highly absorbing peaks (see Sec. \ref{Preparation}).

A ring dye laser (Coherent699-21) using Rhodamine 6G pumped by Nd:YVO$_4$ laser (Coherent Verdi) is used to give 600 mW output power at $\lambda=605.977$ nm. The laser is stabilized against a spectral hole in a second \PrYSO crystal, yielding a coherence time $>$100 $\mu$s and a frequency drift $<$1 kHz/s \cite{Julsgaard2007}. In order to create the desired pulse shapes and to eliminate beam movement accompanying frequency shifts, the laser light was passed twice through a 200 MHz acousto-optic modulator (AOM) with a bandwidth of 100 MHz. A 1 GS/s arbitrary waveform generator (Tektronix AWG520) controlled the AOM, allowing direct control of the light pulse amplitude, phase, and frequency.

After the AOM, the light passed through a single mode optical fiber to clean up the spatial mode. A beam sampler removed a small percentage of the light before the cryostat to be used as a reference beam. The rest of the beam passed through a $ \lambda/2$ plate, such that the light polarization could be aligned along the transition dipole moment direction to give maximum absorption. The beam was then focused to a 100 $\mu$m radius at the center of the sample, which gave Rabi frequencies of maximum 2 MHz for the strongest transitions.

The spectral structures were measured by scanning the light frequency across the spectral structure and recording the intensities of both the transmitted and the reference beams \cite{Rippe2005}. The signals from the detectors were divided to reduce the effect of laser amplitude fluctuations. The intensity of the probe pulses were chosen such that they did not affect the created spectral structures during the readout process. The scan rate was also set such that it had negligible effect on the resolution of the recorded spectra (see discussion below).

\section{Preparation of AFC}
\label{Preparation}
Creating the atomic frequency comb structure, with good control of all necessary parameters, such as peak height, width, separation and number of peaks, can be challenging. Especially considering that the frequency comb structure also preferably should be well separated in frequency from all other absorbing atoms in the material. However, in rare-earth-metal-ion-doped crystals the inhomogeneous absorption profile can indeed be efficiently manipulated, providing the flexibility needed to meet all those requirements. This flexibility is useful not only for AFC, but also for many other similar experiments, such as electromagnetically induced transparency (EIT) \cite{Goldner2009}, slow light \cite{Walther2009a} or quantum computing \cite{Longdell2004,Rippe2008}. Since the precise control of the absorption structure is of particular concern for this paper, we will here make a detailed description of how to create a test platform for AFC experiments and essentially the same technique can be used also for the other listed experiments mentioned above.

\begin{figure*}[ht]
        \includegraphics[width=440pt]{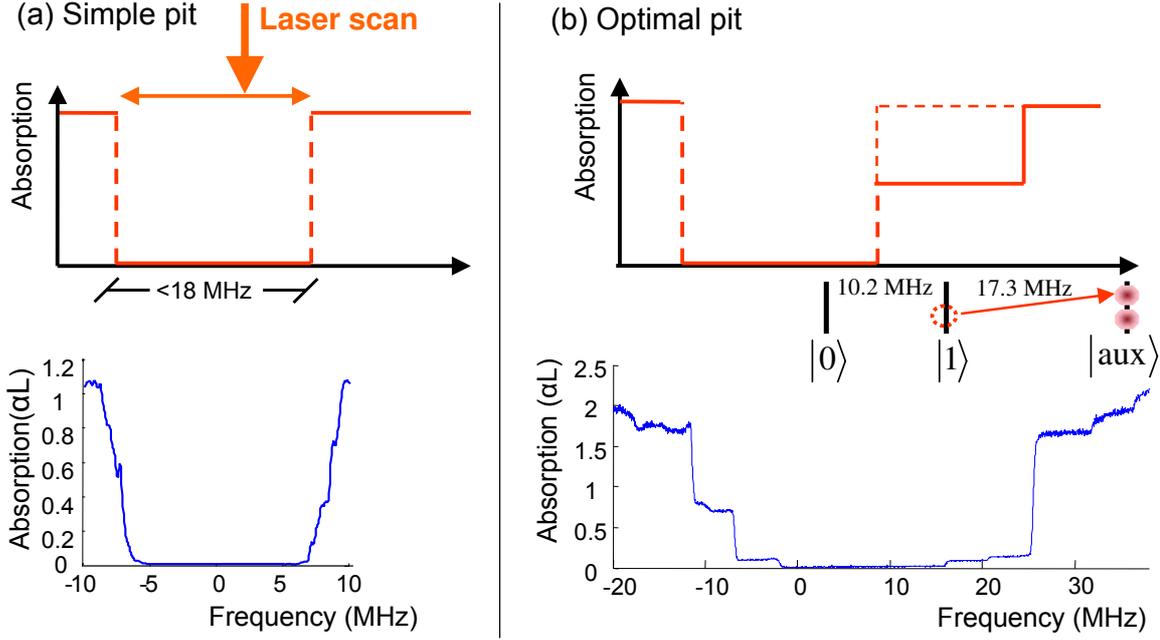}
    \caption{(color online) In a) a simple spectral pit created only by scanning a pulse across a specific interval less than 18 MHz is shown (upper part shows a schematic view and lower part shows actual experimental data). In b) a more optimal pit is shown, where additional burning pulses on different frequency intervals have iteratively been applied to spectrally remove ions as far from the spectral pit as possible (see text). For an exact pulse sequence, see Tables I and II in Appendix. Note that the frequency scale in the two experimental figures is different.}
    \label{Pit_creation}
\end{figure*}

\subsection{Pit creation}

The inhomogeneous $^1D_2-^3H_4$ absorption line in \PrYSO is about 5 GHz and the homogeneous line width of a single Pr ion is about a kHz at temperatures below $\sim$5 K. A chirped laser pulse applied somewhere within the inhomogeneous profile will create a spectral hole through optical pumping. The maximum width of the spectral hole burnt by such a scan is given by the specific level structure of the \Pr3 ion \cite{Equall1995} (see Fig. \ref{Peak_burnback}). After relaxation from excited states the ion has to be in one of the three hyperfine ground levels, which have a total separation of 27.5 MHz, but the maximum spectral hole width is reduced by the hyperfine splitting of the excited state levels of 9.4 MHz, yielding a final spectral hole interval of 18.1 MHz. The scanned pulse will thus create a simple, wide spectral hole, henceforth called a \emph{pit}, and is shown in Fig.~\ref{Pit_creation}a.

When light pulses are applied to perform operations inside the pit, they also have a probability to interact with the tails of the absorption profile of the ions outside the pit, and in particular with the ions immediately outside, forming the walls of the pit. Generally one would like to avoid such interactions and the simple pit in Fig.~\ref{Pit_creation}a then is not optimal. Fortunately, it is possible to shuffle many ions in the walls of the pit further away from the pit. This is illustrated in the top right part of Fig~\ref{Pit_creation}. In this figure, a class of ions having the $\mket{0} \rightarrow \mket{e1}$ transition at some specific frequency just inside the pit and the transitions from the other two ground state levels outside the pit, is displayed. For this ion class, the simple burning pulse only targets the \ket{0}-state so only this ground state will be emptied. However, it is clear that these ions doesn't have to be in the \ket{1} state, in fact, it would be better if they could be further shifted outwards so that all end up in the \ket{aux} state. This can be done by additional burning pulses at the $\mket{1} \rightarrow \mket{e1}$ transitions. This will cause some ions to go to the \ket{aux}-state but also cause some of the ions to fall back down into the pit, and thus, to get the optimal effect, one would have to iterate between the pulses burning at the center of the pit and the ones burning outside it to improve the walls. Similar techniques can be used on the lower frequency side of pit to obtained the final optimal pit, as shown in the lower part of Fig.\ref{Pit_creation}b.

\subsection{Experimental implementation}

The exact sequence of pit burning pulses differ depending on the exact level structure of the ion used. Tab.~\ref{tab:pit_pulses} in the Appendix lists the explicit pulses used to create an optimal pit in \PrYSO in this work, and Table II in the Appendix lists the order in which we have applied those pulses. When working in other materials, essentially the same sequencing can be used, but of course, the actual frequencies have to be changed to match the transitions of the ion in question.

The different optical pumping \emph{BurnPitX} pulses listed in Tab.~\ref{tab:pit_pulses} (see Appendix) are repeated and iterated, as explained in the previous section, in order to create good shallow walls while maintaining no atoms inside the pit. The repetition sequence given at the end of the Appendix is somewhat arbitrary. A higher number of repetitions reflects the fact that the primarily target transition has relatively low oscillator strength. The exact numbers can be changed a bit up or down without significant effect on the result. There is a 1 ms waiting time after every single \emph{BurnPitX} pulse (see Appendix), to give excited ions time to decay back to the ground state before the next pulse arrives. The excited state lifetime is $T_1$=164$\mu$s \cite{Equall1995}.

\subsection{Peak creation}
\label{Peak_creation}
After a suitable pit has been created, a narrow selection of ions is coherently burnt back into the pit. This narrow ensemble of ions now forms an absorption peak spectrally clearly separated from all other ions. The pulses used for this transfer are two complex hyperbolic secant pulses (sechyp for short). The first one targets the $\mket{5/2g} \rightarrow \mket{5/2e}$ transition, of ions having their $\mket{1/2g} \rightarrow \mket{1/2e}$ transition at frequency zero (0) MHz. The second pulse is applied immediately after the first, before the excited ions can decay spontaneously, on the $\mket{5/2e} \rightarrow \mket{1/2g}$ transition. One could imagine taking other routes to burn back a peak, but this particular route is advantageous because the first pulse targets the strongest transition, which means the exciting pulse power, and thus power broadening effects, can be kept at a minimum. The deexcitation pulse on the other hand then targets a weak transition, but since this transition is inside the pit and spectrally far away from other ions, the power can here be increased without any power broadening effects.

\begin{figure}[ht]
    \includegraphics[width=8.5cm]{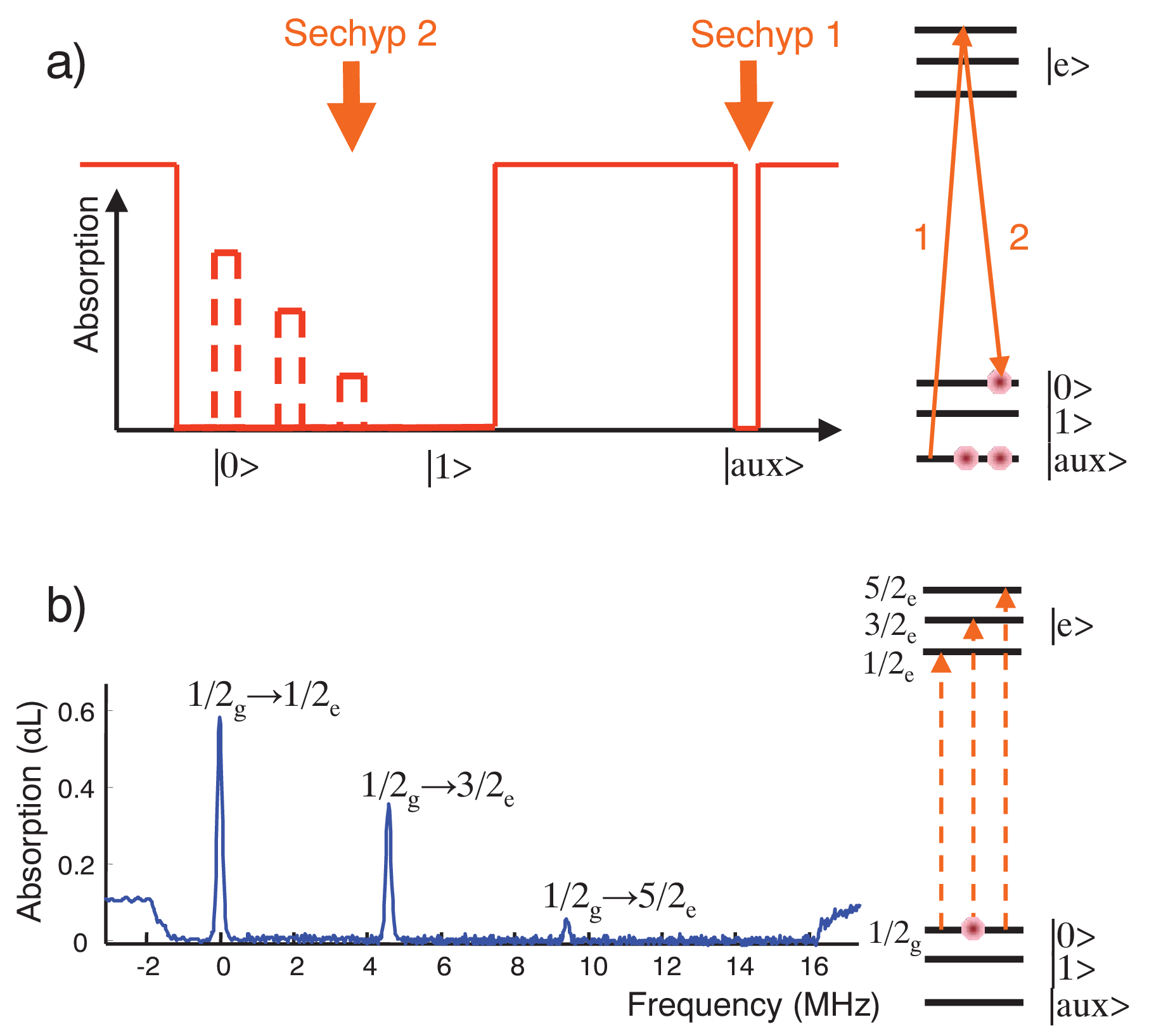}
    \caption{(Color online) Top part shows the sequence and position of the pulses that burns back a narrow ensemble of ions into an empty spectral pit. The lower b) part shows an experimental version of such a created peak structure. The difference in height of the three different transition from the \ket{0} state, comes from the fact that these transitions have different oscillator strength.}
    \label{Peak_burnback}
\end{figure}

Fig.~\ref{Peak_burnback} briefly illustrates how to create such a peak structure, and also displays an experimentally created version. Creating a full atomic frequency comb from this situation, is now the relatively simple matter of adding additional coherent burn-back pulses, with appropriate frequency offsets, creating the additional peaks. The shape of the peaks, as well as width and absorption height, is determined by the burn-back pulses. Changing the spectral shape of these pulses will change the shape of the peaks, and increasing the pulse power will cause more ions to be transferred, which results in higher absorption peaks (as long as there is enough ions available in the crystal at that frequency). This yields a good control over all the essential parameters of the AFC.

\subsection{Comb structure measurement}
One of the goals of this work was to compare the observed AFC echo efficiency with the one predicted by the theoretical model discussed in Sec. \ref{theory}. This requires a precise measurement of the AFC structure in order to determine the shape, width and height of the peaks. To do this the laser was slowly swept in frequency across the created AFC structure and the transmitted light intensity was detected after the sample. From this transmission profile of the structure, the absorption spectrum can be calculated. The intensity of the scan pulse was chosen such that it did not affect the created spectral structure. For most comb measurements we chose to scan the laser over one peak only and lowered the scan rate to a minimum. This is to reduce the effect of the scan rate on the measurement resolution \cite{Chang2005,Rippe2005}. By varying the rate we confirmed that the measured width was indeed independent of the scan rate.

The comb was created on the $\mket{1/2g} \rightarrow \mket{1/2e}$ transition in order to maximize the absorption. It is, however, very challenging to measure absorptions above $d$=3-4. In order to circumvent this problem we instead measured the comb structure on the weaker $\mket{1/2g} \rightarrow \mket{5/2e}$ transition, cf. Fig. \ref{Peak_burnback}. The ratio of these two transitions is known from previous work \cite{Nilsson2004}, thus the optical depth of the $\mket{1/2g} \rightarrow \mket{1/2e}$ transition is readily inferred from the measured absorption spectrum.

\section{Results and discussion}
\label{result}

\begin{figure}[ht]
     \includegraphics[width=.50\textwidth]{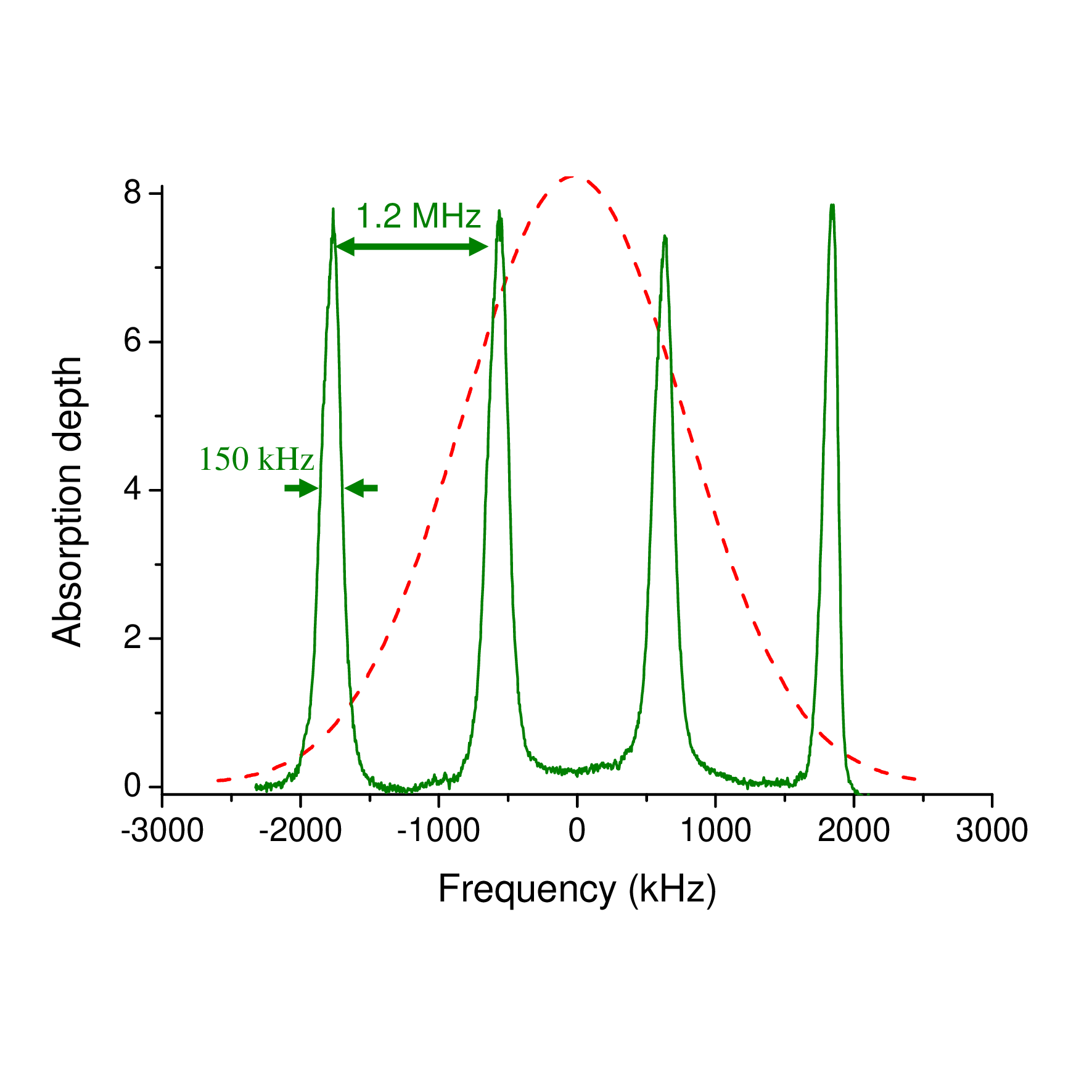}
    \caption{(Color online) From the inhomogeneous absorption profile, four peaks with ions, all absorbing on the $\mket{1/2g} \rightarrow \mket{1/2e}$ transition are created. The peak width (FWHM) is $\gamma = 150$ kHz and they are separated by $\Delta=1.2$ MHz. The input pulse has a Gaussian power spectrum with FWHM=2 MHz.}
    \label{AFC_comb}
\end{figure}

The input pulse is stored on the $\mket{1/2g} \rightarrow \mket{1/2e}$ transition, which is the transition for ions in state $\mket{1/2g}$ with the highest oscillator strength. This results in a comb with high optical depth $d$. The bandwidth of the AFC is limited by the frequency separation between the excited states and in the present case this is about $4.6$ MHz, as set by the $\mket{1/2e}$ and $\mket{3/2e}$ separation (cf. Fig. \ref{Peak_burnback}). Fig~\ref{AFC_comb}, shows a comb containing four peaks, and where the width of each peak is about $\gamma$= 150 kHz. The separation between the peaks was set to, $\Delta$=1.2 MHz. The pulse to be stored has a Gaussian shape with duration 200 ns, resulting in a frequency power spectrum with FWHM 2 MHz (see Fig. \ref{AFC_comb}).

\begin{figure}[ht]
    \includegraphics[width=.50\textwidth]{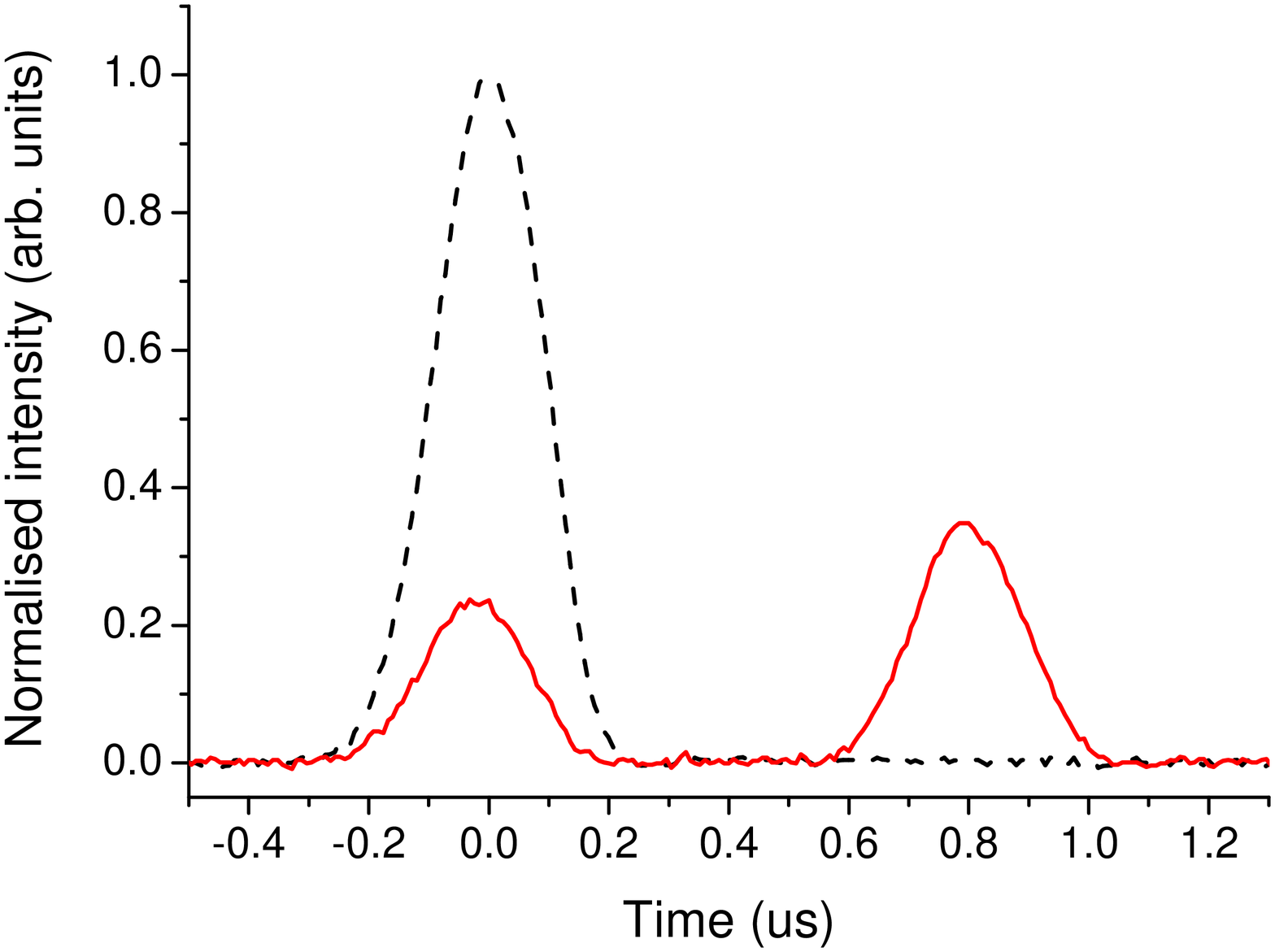}
    \caption{(Color online) The dashed line shows the input pulse that is completely transmitted through the empty pit (no absorption). The solid line shows the partially transmitted input pulse and the subsequent echo emission with the AFC created in the pit (cf. Fig \ref{AFC_comb}). The echo efficiency is 35\% of the input pulse (see text for details).}
    \label{echo_eff_34}
\end{figure}

A high-efficiency echo is shown in Fig. \ref{echo_eff_34}. The emitted echo is observed after 800 ns, as expected from the comb periodicity $\tau_s=1/\Delta$=800ns. To be able to calculate the efficiency of the echo, first a reference input pulse is first sent through the empty pit (no AFC prepared). This pulse is thus completely transmitted through the sample. The AFC is then prepared inside the pit and an identical pulse (the storage pulse) is sent in. This pulse is partially absorbed in the medium and produces an echo. The ratio between the area of the echo and the area of the reference pulse gives the storage efficiency. A small part of the reference pulse as well as of the storage pulse is split off before they enter the cryostat so unintentional input power differences between the reference and storage pulses can be compensated for. For the data shown in Fig. \ref{echo_eff_34} we measured an efficiency of 35\%. To our knowledge this is the highest AFC echo efficiency observed up to date.

As discussed in Sec. \ref{theory} the efficiency depends strongly on the shape of the AFC. Our comb structure measurements show peaks that have near Gaussian shapes. This facilitates the theoretical modeling since we can use the simple model discussed in Sec. \ref{theory} and we need only then to measure two parameters; the peak absorption $d$ and peak width $\gamma$ (the finesse is calculated from the relationship $F=\Delta/\gamma$). In order to make a quantitative comparison with the model, we varied the peak absorption $d$ and measured the resulting input pulse transmission and echo efficiency. This was done by increasing the power of the back burning pulses used in the peak creation (see Sec. \ref{Peak_creation}). For each power setting we also measured the comb structure to find $d$ and $F$. The width of the peaks could be varied by changing the chirp width of the sechyp pulses used for peak creation. We did measurements as a function of $d$ for two settings of the chirp; 200 and 300 kHz. The measured peak widths for these two settings were 175 and 245 kHz, respectively, corresponding to $F$=6.9 and 4.9. With increasing back burning power (hence increasing $d$) the peaks were slightly broadened due to power broadening, but the observed increase was only 10-15\% for the data considered here. The widths given above are averages over all back burning powers (hence $d$ values).

\begin{figure}[ht]
     \includegraphics[width=.50\textwidth]{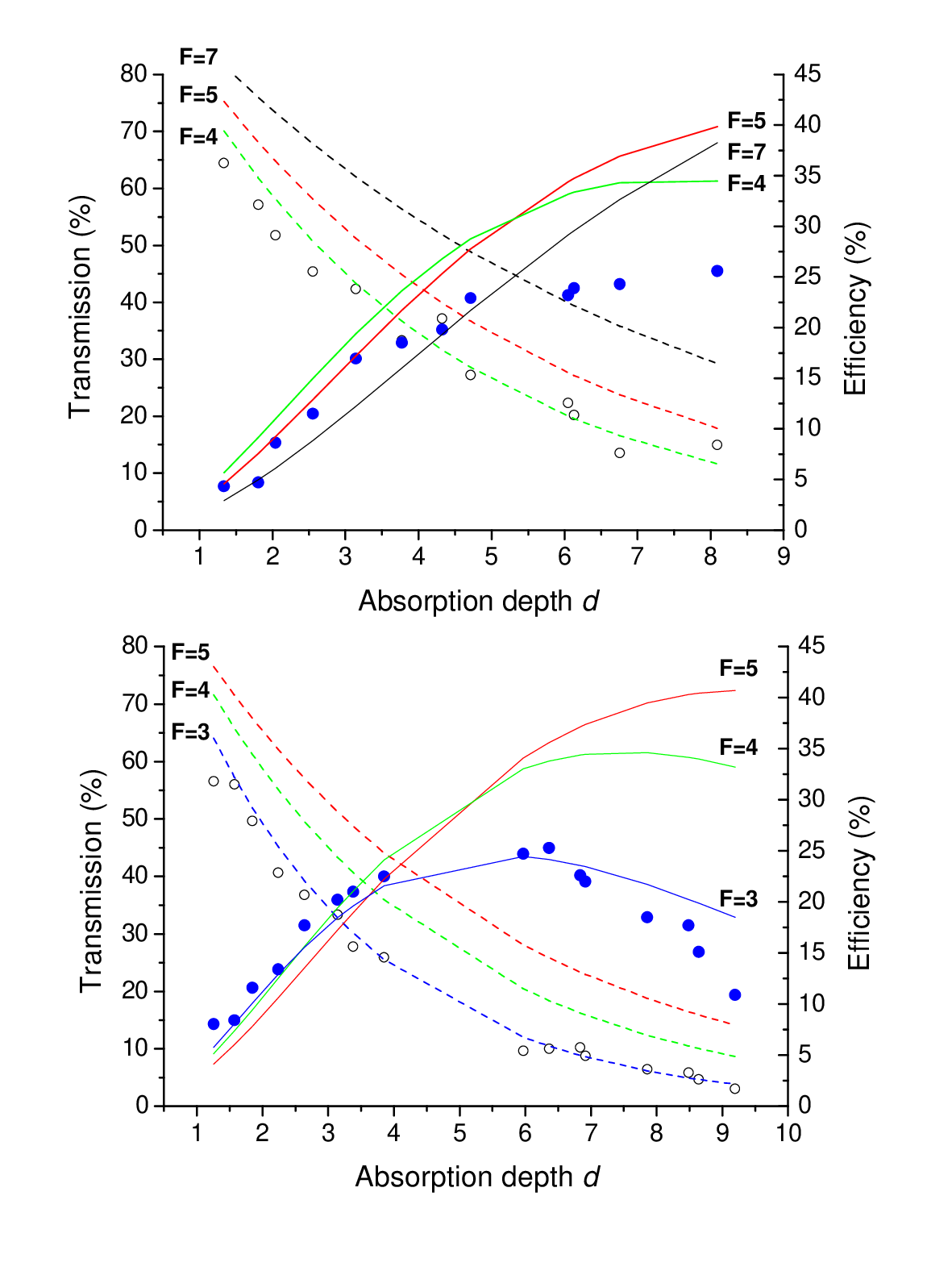}
    \caption{(Color online) Measured transmission of the input pulse (open circles and left axis) and echo efficiency (closed circles and right axis) as a function of the measured optical depth for two different experimental values of the finesse, (a) $F=6.9$ and (b) $F=4.9$. The dashed lines are theoretically calculated transmission coefficients and solid lines are calculated efficiencies for (a) F=4,5 and 7, and for (b) F=3,4 and 5 (see text for details).}
\label{300GF345}
\end{figure}

In Fig. \ref{300GF345} we show measured transmission coefficients of the input pulse and the efficiencies of the echo for the two data sets. The data is plotted as a function of the measured peak absorption $d$ as extracted from the AFC spectra. Theoretical transmission and efficiency curves are also shown. These were calculated using the experimental $d$ values, for different values of finesse. It is observed that the transmission coefficient is very sensitive to the finesse, whereas the efficiency is less sensitive up to $d$=4-5. In general the best agreement for the transmission is obtained for a finesse lower than the one measured from the comb spectra (see above). We can also see that the best-fit finesse is lower for the 300 kHz data set, than for the 200 kHz, which is to be expected. The echo efficiency shows a reasonably good agreement with all three values of the finesse up to $d$=4-5. But for $d \geq 5$ the discrepancy between the experimental and theoretical values becomes significant for $F$=4 and 5, while for $F$=3 it is still satisfactory.

The general trend, for both transmission and efficiency, is that our data fit better with a finesse lower than the measured one. This comparison is however made within the theoretical framework of Sec. \ref{theory} where a comb of Gaussian peaks was assumed. Both the transmission and echo efficiency are strongly dependent on the actual peak shape \cite{Afzelius2009a,Chaneliere2009}. For instance, the same comparison with a Lorentzian model \cite{Chaneliere2009} yields very different best-fit values for the finesse. Although much effort was devoted to the precise measurement of the comb structure, it is still conceivable that the actual peak shape deviate from pure a Gaussian shape. Particularly since the sechyp pulses used for peak creation have power spectra with more super-Gaussian shape \cite{Rippe2005}. Another source of error could be imperfections in the peak creation pulses, where the high power needed to obtain high optical depth might generate an increase in the absorption background due to off-resonant excitation. Such an additional absorption will reduce both the experimental transmission and efficiency \cite{Riedmatten2008} compared to the theoretical model in Sec. \ref{theory} where such an absorption background is neglected. This would particularly affect the high $d$ range, where indeed we observe a larger discrepancy.

\section{Conclusions}
\label{conclus}

We have in detail described optical pumping and preparation procedures for creating AFC structures in \PrYSO. We were able to make comb structures yielding $>30$\% AFC echo efficiency, which are the most efficient AFC echoes observed up to date. We believe that further progress will be possible, by carefully optimizing the comb parameters. It should thus be possible to approach the theoretical limit of 54\% used in the present forward-propagation configuration. In order to make a significant further progress, recall in the backward direction would be necessary, in which case 100\% efficiency is theoretically possible in the absence of dephasing.

In this work we also compared the experimentally observed efficiencies to a theoretical model. Considerable care has been put into determining the line shape and line width of the generated AFC structure with good precision in order to be able to theoretically model the experimental efficiencies. Still, at high optical densities the finesse required to theoretically reproduce the experimental echo efficiency are lower than those measured experimentally. Nevertheless, the present results indeed show that high efficiency QMs can be created using the AFC technique.

\section{Acknowledgements}

This work was supported by the Swedish Research Council, the Knut and Alice Wallenberg Foundation, the Swiss NCCR
Quantum Photonics, the European Commission through the integrated project QAP, and the ERC Advanced Grant QORE.

\section{APPENDIX}	

The optical pumping pulses used for creating the pit structure in Fig.~\ref{Pit_creation}b are shown in Tab.~\ref{tab:pit_pulses}.

\subsubsection{Table I}

\begin{table}[h]
\begin{tabular}{lrrl}
  \emph{Pulse} & $\nu_{start}$/MHz & $\nu_{end}$/MHz & $\Omega_{rel}$ \\ \hline
  BurnPit1 & +31.85 & +24.15 & $3/2_g \rightarrow 1/2_e$ \\
  BurnPit2 & +23.85 & +16.15 & $3/2_g \rightarrow 5/2_e$ \\
  BurnPit3 & +15.95 & +7.65 & $3/2_g \rightarrow 5/2_e$ \\
  BurnPit4 & +23.85 & +16.15 & $3/2_g \rightarrow 5/2_e$ \\
  BurnPit5 & -16.85 & -9.15 & $5/2_g \rightarrow 5/2_e$ \\
  BurnPit6 & -8.85 & -1.15 & $5/2_g \rightarrow 1/2_e$ \\
  BurnPit7 & +15.95 & +7.65 & $3/2_g \rightarrow 5/2_e$ \\
  BurnPit8 & +7.35 & -1.10 & $3/2_g \rightarrow 5/2_e$ \\
  BurnPit9 & -1.10 & +7.35 & $5/2_g \rightarrow 1/2_e$ \\
  BurnPit10 & +7.65 & +15.95 & $5/2_g \rightarrow 1/2_e$
\end{tabular}
\caption{List of pulses used for the pit burning sequence, with start and end frequencies. This set of pulses will create the pit Fig.~\ref{Pit_creation}b with zero absorption (no absorbing ions) from -1.2 MHz up to 16.2 MHz. The frequency scale is defined by denoting the $\mket{1/2g} \rightarrow \mket{1/2e}$ transition, for an arbitrarily selected ion class, as zero MHz. The column $\Omega_{rel}$ lists the primary target transition of the scan for the purpose of knowing what light intensity to choose in  order to match the Rabi frequency to the relative oscillator strength. Note that pulses number 2 and 4 are the same, as is pulses 3 and 7.}
\label{tab:pit_pulses}
\end{table}

These pulses are then repeated in an iterative sequence in the following manner

\begin{enumerate}
  \item \emph{Repeat 60 times:} BurnPit5, BurnPit6
  \item \emph{Repeat 30 times:} BurnPit1-4, BurnPit6-10
  \item \emph{Repeat 20 times:} BurnPit1-4, BurnPit6
  \item \emph{Repeat 30 times:} BurnPit7-10
\end{enumerate}
	
finally yielding the pit in Fig.~\ref{Pit_creation}b.

\section{References}
\bibliographystyle{elsarticle-num}

\begin{thebibliography}{10}
\expandafter\ifx\csname url\endcsname\relax
  \def\url#1{\texttt{#1}}\fi
\expandafter\ifx\csname urlprefix\endcsname\relax\def\urlprefix{URL }\fi
\expandafter\ifx\csname href\endcsname\relax
  \def\href#1#2{#2} \def\path#1{#1}\fi

\bibitem{Briegel1998}
H.-J. Briegel, W.~D\"{u}r, J.~I. Cirac, P.~Zoller, Phys. Rev. Lett. 81
  (1998) 5932.

\bibitem{Duan2001}
L.~M. Duan, M.~D. Lukin, J.~I. Cirac, P.~Zoller, Nature 414 (2001) 413.

\bibitem{Simon2007}
C.~Simon, H.~de~Riedmatten, M.~Afzelius, N.~Sangouard, Z.~Zbinden, N.~Gisin, Phys. Rev.
  Lett 98 (2007) 190503.

\bibitem{Sangouard2009}
N.~Sangouard, C.~Simon, H.~de~Riedmatten, N.~Gisin, \emph{Quantum repeaters based on atomic ensembles and linear optics}, arXiv:0906.2699.

\bibitem{Julsgaard2004a}
B.~Julsgaard, J.~Sherson, J.~I. Cirac, J.~Fiurasek, E.~S. Polzik, Nature 432 (2004) 482.

\bibitem{Chaneliere2005}
T.~Chaneli\`{e}re, D.~N. Matsukevich, S.~D. Jenkins, S.-Y. Lan, T.~B. Kennedy, Nature 438 (2005) 833.

\bibitem{Eisaman2005}
M.~D. Eisaman, A.~Andre, F.~Massou, M.~Fleischhauer, A.~S. Zibrov, M.~D. Lukin, Nature 67 (2005) 452.

\bibitem{Choi2008}
K.~S. Choi, H.~Deng, J.~Laurat, H.~J. Kimble, Nature 452 (2008) 67.

\bibitem{Riedmatten2008}
H.~de~Riedmatten, M.~Afzelius, M.~U. Staudt, C.~Simon, N.~Gisin, Nature 456 (2008) 773.

\bibitem{Collins2007}
O.~A. Collins, S.~D. Jenkins, A.~Kuzmich, T.~A.~B. Kennedy, Phys. Rev. Lett. 98 (2007) 060502.

\bibitem{Hammerer2008}
K.~Hammerer, A.~Sorensen, E.~Polzik, \emph{Quantum interface between light and atomic
  ensembles}, arXiv:0807.3358.

\bibitem{Ruggiero2009}
J.~Ruggiero, J.~L. Le~Gou\"{e}t, C.~Simon, T.~Chaneli\`{e}re, Phys. Rev. A 79 (2009)
  053851.

\bibitem{Moiseev2001}
S.~A. Moiseev, S.~Kr\"oll, Phys. Rev. Lett. 87 (2001) 173601.

\bibitem{Nilsson2005a}
M.~Nilsson, S.~Kr\"{o}ll, Opt. Comm. 247 (2005) 393.

\bibitem{Kraus2006}
B.~Kraus, W.~Tittel, N.~Gisin, M.~Nilsson, S.~Kr\"{o}ll, J.~I. Cirac, Phys. Rev. A 73 (2006) 020302(R).

\bibitem{Alexander2006}
A.~L. Alexander, J.~J. Longdell, M.~J. Sellars, N.~B. Manson, Phys. Rev. Lett. 96 (2006) 043602.

\bibitem{Tittel2008}
W.~Tittel, M.~Afzelius, R.~L. Cone, T.~Chaneli\`{e}re, S.~Kr\"{o}ll, S.~A.
  Moiseev, M.~Sellars, \emph{Photon-echo quantum memory}, arXiv:0810.0172.

\bibitem{Afzelius2009a}
M.~Afzelius, C.~Simon, H.~Riedmatten, N.~Gisin, Phys. Rev. A 79 (2009) 052329.

\bibitem{Staudt2007a}
M.~U. Staudt, S.~R. Hastings-Simon, M.~Nilsson, M.~Afzelius, V.~Scarani,
  R.~Ricken, H.~Suche, W.~Sohler, W.~Tittel, N.~Gisin, Phys. Rev. Lett. 98 (2007)
  113601.

\bibitem{Chaneliere2009}
T.~Chaneli\'{e}re, J.~Ruggiero, M.~Bonarota, M.~Afzelius, J.~L. Le~Gou\"{e}t,
  \emph{Efficient light storage in a crystal using an atomic frequency comb},
  arXiv:0902.2048.

\bibitem{Usmani2009}
I.~Usmani, M.~Afzelius, H.~de~Riedmatten, N.~Gisin, \emph{Mapping tens of photonic
  qubits onto one solid-state atomic ensemble}, in preparation.

\bibitem{Macfarlane1987}
R.~M. Macfarlane, R.~M. Shelby, Coherent transient and holeburning spectroscopy
  of rare earth ions in solids, in: A.~Kaplyanskii, R.~Macfarlane (Eds.),
  Modern problems in condensed matter sciences, North-Holland, Amsterdam, 1987.

\bibitem{Macfarlane2002}
R.~M. Macfarlane, J.Lumin. 100 (2002) 1.

\bibitem{Hesselink1979}
W.~H. Hesselink, D.~A. Wiersma, Phys. Rev. Lett. 43 (1979) 1991.

\bibitem{Carlson1984}
N.~W. Carlson, Y.~S. Bai, W.~R. Babbitt, T.~W. Mossberg, Phys. Rev. A 30 (1984) 1572.

\bibitem{Mitsunaga1991a}
M.~Mitsunaga, R.~Yano, N.~Uesugi, Opt.Lett. 16 (1991) 264.

\bibitem{Yano1992}
R.~Yano, M.~Mitsunaga, N.~Uesugi, J. Opt. Soc. Am. B 9 (1992) 992.

\bibitem{Merkel1996b}
K.~D. Merkel, W.~R. Babbitt, Opt. Lett. 21 (1996) 1102.

\bibitem{Tian2001a}
M.~Tian, J.~Zhao, Z.~Cole, R.~Reibel, W.~R. Babbitt, J. Opt. Soc. Am. B 18 (2001) 673.

\bibitem{Afzelius2009b}
M.~Afzelius, I.~Usmani, A.~Amari, B.~Lauritzen, A.~Walther, C.~Simon,
  N.~Sangouard, J.~Minar, H.~de~Riedmatten, N.~Gisin, S.~Kr\"oll, \emph{Demonstration
  of atomic frequency comb memory for light with spin-wave storage},
  arXiv:0908.2309.

\bibitem{Nilsson2005}
M.~Nilsson, L.~Rippe, S.~Kr\"{o}ll, R.~Klieber, D.~Suter, Phys. Rev. B 70 (2004) 214116.

\bibitem{Sangouard2007}
N.~Sangouard, C.~Simon, M.~Afzelius, N.~Gisin, Phys. Rev. A 75 (2007) 032327.

\bibitem{Walther2009a}
A.~Walther, A.~Amari, A.~Kalachev, S.~Kr\"{o}ll, Phys. Rev. A 80 (2009) 012317.

\bibitem{Julsgaard2007}
B.~Julsgaard, L.~Rippe, A.~Walther, S.~Kr\"{o}ll, Optics Express 15 (2007) 11444.

\bibitem{Rippe2005}
L.~Rippe, M.~Nilsson, R.~Klieber, D.~Suter, S.~{Kr\"oll}, Phys. Rev. A 71 (2005) 062328.

\bibitem{Goldner2009}
P.~Goldner, O.~Guillot-No\"{e}l, F.~Beaudoux, Y.~L. Du, J.~Lejay, J.~L.
  Le~Gou\"{e}t, T.~Chaneli\`{e}re, L.~Rippe, A.~Amari, A.~Walther,
  S.~Kr\"{o}ll, Phys. Rev. A 79 (2009) 033809.

\bibitem{Longdell2004}
J.~J. Longdell, M.~J. Sellars, Phys. Rev. A 69 (2004) 032307.

\bibitem{Rippe2008}
L.~Rippe, B.~Julsgaard, A.~Walther, Y.~Ying, S.~Kr\"{o}ll, Phys. Rev. A 77 (2008) 022307.

\bibitem{Equall1995}
R.~W. Equall, R.~L. Cone, R.~M. Macfarlane, Phys. Rev. B 52 (1995) 3963.

\bibitem{Chang2005}
T.~Chang, M.~Z. Tian, R.~K. Mohan, C.~Renner, K.~D. Merkel, W.~R. Babbitt, Opt. Lett. 30 (2005) 1129.

\bibitem{Nilsson2004}
M.~Nilsson, L.~Rippe, R.~Klieber, D.~Suter, S.~{Kr\"oll}, Phys. Rev. B 70 (2004) 214116.

\end{thebibliography}

\end{document}